\documentclass{iaus}
\usepackage{graphicx}

\title[Pre-SN Mass Loss] 
{Episodic Mass Loss and Pre-SN Circumstellar Envelopes}
\author[Nathan Smith]{Nathan Smith}
\affiliation{Astronomy Department, University of California, Berkely,
CA 94720, USA \\ email: {\tt nathans@astro.berkeley.edu}}

\pubyear{2008}
\volume{250}  
\pagerange{xxx-xxx}
\jname{Massive Stars as Cosmic Engines}
\editors{F.\ Bresolin, P.A.\ Crowther, \& J.\ Puls, eds.}
\begin{document}
\maketitle

\begin{abstract}
I discuss observational clues concerning episodic mass-loss properties
of massive stars in the time before the final supernova explosion.  In
particular, I will focus on the mounting evidence that LBVs and
related stars are candidates for supernova progenitors, even though
current paradigms place them at the end of core-H burning.  Namely,
conditions in the immediate circumstellar environment within a few
10$^2$ AU of Type IIn supernovae require very high progenitor
mass-loss rates.  Those rates are so high that the only known stars
that come close are LBVs during rare giant eruptions.  I will
highlight evidence from observations of some recent extraordinary
supernovae suggesting that explosive or episodic mass loss (a.k.a.\
LBV eruptions like the 19th century eruption of Eta Car) occur in the
5-10 years immediately preceding the SN.  Finally, I will discuss some
implications for stellar evolution from these SNe, the most important
of which is the observational fact that the most massive stars can
indeed make it to the ends of their lives with substantial H envelopes
intact, even at Solar metallicity.
\end{abstract}

\firstsection 
\section{Introduction}

Supernovae are one of the most influential ways that massive stars act
as cosmic engines, energizing and polluting the ISM (especially the
early ISM).  To evaluate the role of massive stars as cosmic engines
observationally, we must first understand the relationship betwen
massive star evolution (mass loss, rotation, metallicity) and the type
of SN observed, since SNe and can be seen to large distances.  The
flip side to that coin is that if we can understand the relationship
between different SN properties and the star's evolution locally, then
SNe also become our best probe of mass loss, circumstellar structure,
stellar evolution, the initial mass function, and star formation rates
throughout cosmic time.  In fact, one could argue that SNe are the
{\it only} reliable way to directly probe mass loss of individual
stars at large distances where spatially resolving individual stars
and separating their light from their host environment is hopeless.

One of the most important questions concerns the connection between
the type of SN observed and the star's intial mass.  Whether a SN is a
normal Type II with H (8--20 $M_{\odot}$ stars), a Type Ib/c that has
shed its H envelope and probably died as a WR star, or a Type IIn with
dense circumstellar material, is determined mainly by the star's
mass-loss rate during evolution.  A key prediction of most current
stellar evolution models is that all stars above some mass, say
$\sim$30 $M_{\odot}$, will fully shed their H envelopes and explode as
WR stars, making SNe of Type Ib/c.  Recent observations of SNe, on the
other hand, are making it difficult to avoid the conclusion that this
prediction is wrong.

Namely, SNe that have dense circumstellar material suggest to us that
a small fraction of stars --- probably the most massive stars --- have
violent mass-loss events that precede the final SN explosion.  When H
is present, these are seen as Type IIn SNe, making the most luminous
SNe in the Universe.

I should preface the discussion below by saying that exploding LBVs or
LBV-like stars represent a small fraction of observed core-collapse SN
events.  The vast majority are normal Type II's (see Smartt, these
proceedings), while 15--20\% are normal Type Ib/c SNe.  That is
perhaps not so surprising though, given that the most massive stars
and LBVs themselves are quite rare.

\section{LBVs as Supernova Progenitors}

I'll start with a list of observational clues that some SNe occur when
their progenitor star was in (or had recently been in) the LBV phase,
contrary to expectations.  Each of these on their own is compelling if
not necessarily convincing, but taken together, they paint a
consistent picture of LBVs as SN progenitors that is hard to dismiss.

\begin{itemize}

\item{\bf SN 2006jc:} While this was not a Type IIn event (it was a
peculiar Type Ib), it is unique so far in that it is the only SN that
was actually {\it observed} to have an LBV-like outburst 2 yr before
exploding as a SN.  The eventual supernova showed an unusual spectrum
and dust formation that was caused by very dense circumstellar
material (see Foley et al.\ 2007; Pastorello et al.\ 2007; Smith et
al.\ 2008).

\item{\bf SN 2005gl:} This Type IIn supernova had a progenitor star
identified on pre-explosion images, which was a blue/yellow supergiant
that had a luminosity and colors indicative of an LBV star (Gal-Yam et
al.\ 2007).

\item{\bf Radio modulations:} Kotak \& Vink (2006; and these
proceedings) suggested that semi-periodic modulations in the radio
lightcurves of a few SNe might be caused by the shock running into
density variations caused by normal S Dor-type excursions of an LBV
progenitor.  The suggested cases were SNe~2001ig and 2003bg, both of
which transitioned between Type Ib/c and Type II spectra (or the
reverse) as they evolved, and possibly also SN~1979C and 1998bw.
However, I should note that alternative interpretations of the radio
modulations have been forwarded as well (see Ryder et al.\ 2004;
Soderberg et al.\ 2006; Schwarz \& Pringle 1996).

\item{\bf Circumsellar Nebulae:} There are three objects in our galaxy
that are near twins of the unusual ring nebula around SN~1987A:
HD~168625 is an LBV (Smith 2007), while SBW1 (Smith et al.\ 2007a) and
Sher 25 (Smart et al.\ 2002) have abundances inconsistent with passage
through a previous RSG phase.  The nearly identical nebular ring
structures suggest that the progenitor of SN~1987A had recently been
in a similar evolutionary state when it exploded (see Smith 2007 and
Smith et al.\ 2007a for further discussion).

\item{\bf Luminous Type IIn Supernovae:} The very high-luminosity of
some Type IIn SNe can only be accounted for if the star had a huge
mass-loss event in the decade before te SN.  The clearest cases, which
are also the three most luminous SNe ever observed, are SN~2006gy,
SN~2005ap, and 2006tf (Smith et al.\ 2007b; Ofek et al.\ 2007; Smith
\& McCray 2007; Woosley et al.\ 2007; Quimby et al.\ 2007, and Smith
et al., in prep.); these require mass-loss rates on the order of 1
$M_{\odot}$/yr in the few years before the SN.  Other Type IIn's, like
SN 1979C, 1988Z, and others, also suggest LBV-like mass loss based on
their extended high-luminosity CSM interaction.  In some cases, like
SN2006gy and 2006tf, the progenitor's wind speed is {\it observed} in
the narrow P Cyg H$\alpha$ absorption, and its speed is $\sim$200
km/s.  That's too fast for a RSG, but just right for an LBV.

\end{itemize}

This last point is probably the most intersting, in my view, because
it highlights a relatively unfamiliar phenomenon that is truly
remarkable, and that I think deserves more attention.  Namely, the
high luminosity of these Type IIn SNe require huge bursts of episodic
mass loss {\it right before they explode}.  In order to produce a Type
IIn spectrum, the luminosity from CSM interaction must be comparable
to or larger than the luminosity from recombination of the SN ejecta
or radioactive decay, which are characteristically about 10$^9$
$L_{\odot}$.  A convenient expression for the progenitor's mass-loss
rate needed to produce an observed luminosity $L_9$=$L_{SN}/(10^9
L_{\odot})$ through CSM interaction, with an optimistic 100\%
efficiency of converting shock kinetic energy into visual light, is
given by

\begin{displaymath}
\dot{M} = 0.04 \ L_9 \frac{v_{w}}{200} \big{(} \frac{v_{SN}}{4000} \big{)}^{-3} M_{\odot
} {\rm yr}^{-1}
\end{displaymath}

\noindent where $v_{w}$ and $v_{SN}$ are the progenitor's wind speed
and the SN blast wave speed, respectively, in km s$^{-1}$.  L can be
measured from the light curve, while $v_{w}$ and $v_{SN}$ can usually
be measured from the narrow and relatively broad components of the
H$\alpha$ line.  For the main peak of SN~2006gy, the observed
luminosity of $L_9$=50 at 70 d after explosion and the observed speeds
of $v_{w}\simeq$200 km/s and $v_{SN}\simeq$4,000 km/s required a
mass-loss rate for the progenitor of $\sim$2 $M_{\odot}$ yr$^{-1}$ for
5--10 yr.  That's the most extreme example, but its easy to see that
mass-loss rates more than about 10$^{-2}$ $M_{\odot}$/yr are needed in
order for the CSM interaction luminsity to compete with the normal
luminosity source of the SN.

If we look around us and ask ``Among known stars in the Universe,
which ones have the requisite mass loss to produce Type IIn SNe?'',
the only viable cadidates with mass-loss rates above 10$^{-2}$
$M_{\odot}$/yr are {\it LBVs during a giant eruption}.  If they are
not bona-fide LBVs, then Type IIn progenitors are doing a darn good
job of impersonating the H composition, wind speeds, and mass-loss
rates of LBVs.

\section{Synchronicity}

The argument that the heavy mass loss occurs in the decade or so
immediately preceding the SN is pretty straightforward.  In the 100
days or so after explosion when the SN is bright and shows a Type IIn
spectrum, it is sweeping through a radius of only a few 100 AU
(typical observed blast wave speeds are only a few 1000 km/s, because
the dense material decelerates the blast wave).  In Type IIn SNe, the
progenitor's wind speed can be observed in the narrow P Cygni
absorption of H$\alpha$, usually indicating speeds of a couple 100
km/s.  Thus, the radius out to which the blast wave reaches in the
time it is being observed corresponds to the star's mass loss during
the previous decade.

\section{Deaths of the Most Massive Stars}

Type IIn supernovae give us the most luminous SNe known in the
Universe.  A natural tendency is to associate them with the deaths of
the most massive stars.  Here are some reasons to favor that
interpretation:

The classical LBVs that are expected to have giant eruptions {\it a
la} Eta Carinae are the most luminous stars known, with initial masses
of 60--150 $M_{\odot}$, although there are also some lower luminosity
LBVs that are probably post-RSGs (Smith, Vink, \& de Koter 2004) which
may arise from stars with initial masses of perhaps 25--40
$M_{\odot}$.  Since LBV eruptions provide our only observed precedent
for the required mass-loss rates of Type IIn SNe, the simplest
assumption is that Type IIn's do indeed represent the rare deaths of
the most massive stars.

Aside from giant LBV eruptions, the ``pulsational pair instability''
is the best bet going (Woosley et al.\ 2007), as it is the only
theoretically-proposed mechanism to produce mass loss similar to LBV
eruptions, and it is expected to occur right before the SN.  However,
the instability only occurs for initial mass above $\sim$95 (Woosley
et al.\ 2007).  Thus, if this mechanism is responsible for the pre-SN
mass loss of Type IIn's, then they {\it REALLY} must be the deaths of
the most massive stars.

The brightest Type IIn SNe are energetic events, with combined
luminous + kinetic energies well in excess of 10$^{51}$ ergs (for
example, 06gy emitted more than that in visible light alone).
High-energy SN explosions are not something we associate with stars of
moderate mass (i.e. 8-20 $M_{\odot}$).

(This is a bit of a tangent, but recent results show that $\sim$30
$M_{\odot}$ black holes exist [Prestwich et al.\ 2007; Silverman \&
Filippenko 2008, in prep.].  Those must come from the core-collapse
deaths of very massive stars that did shed all their H, because the
most massive H-free WR stars are less than that [Smith \& Conti
2008].)

The luminous Type IIn's, like SN~2006gy, appear to eject 10's of
$M_{\odot}$ in the decade or so before the SN.  Very massive stars
seem to be able to do this (Smith \& Owocki 2006), but it is hard to
believe that an 8--20 $M_{\odot}$ star could shed that much of its
mass in a couple years.  Furthermore, we would have no explanation for
why only a small fraction of these stars have violent precursor
events, whereas most die as normal Type II-P SNe.  On the other hand,
the most massive stars are rare compared to stars of 8--40
$M_{\odot}$, so its natural that their deaths would be a small
fraction of all core-collapse SNe.

Lastly, wind speeds of progenitor stars can be gleaned from the narrow
P Cygni H$\alpha$ absorption in the spectra of Type IIn SNe, and they
typically have fast winds of a few 10$^2$ km/s, characteristic of LBVs
(see Smith et al.\ 20076b).  Moderately massive stars (initial mass
8--20 $M_{\odot}$) should die as RSGs, with wind speeds of 10--20
km/s.

{\underline{\it Conclusion:}} Type IIn progenitors are NOT moderately
massive stars (initial masses of 8-20 $M_{\odot}$), but must be very
massive stars, with initial masses that are probably above 50--60
$M_{\odot}$.  Since they die with a lot of H, this is bad news for
stellar evolution models.

\section{LBVs, LBV Impostors, SN Impostors, and SNe}

Since LBV-like eruptions are apparently responsible for the conditions
that make the most luminous SNe in the Universe, our ignorance of
their underlying physical mechanism is rather embarassing.


1.  One possibility is that the progenitor stars are in a regular LBV
phase, that giant LBV eruptions occur repeatedly, and that some of
these coincidentally occur shortly before the final SN explosion.  The
natural implication is that while some eruptions occur within a decade
or so of the SN, many more probably will not.  Statistically, this is
a bit troubling, because Type IIn's represent 2--5\% of core
collapse SNe (Capallaro et al.\ 1997).  If those correspond to the
ones that have had giant eruption-like mass loss in the decades before
the SN, then there must be at least 10 times more that are in a
quiescent phase between giant eruptions. To achieve this from a normal
Salpeter IMF, we would need to have {\it all} stars with initial
masses above about 40 $M_{\odot}$ die as LBVs.


2.  A second possibility is that these LBV-like eruptions really
    represent a {\it precursor} to the Type IIn SN, and their
    synchronization is not a coincidence.  This could be the case if
    the SN-precursor mass ejections are in fact associated with an
    instability, perhaps the pulsational pair instability or some
    other instability leading to explosive mass loss in the very final
    nuclear burning stages.  In that case, to get the observed rate of
    Type IIn's from a Salpeter IMF, we'd need all stars above 85--90
    $M_{\odot}$ to explode in this way.  Interestingly, this is also
    the range of masses that are supposed to be susceptible to the
    pulsational pair instability (Woosley et al.\ 2007).


In either case, how would we know?  Can we tell the difference
observationally between a classical LBV and a pulsational pair event?
It is important to reiterate that regardless of the underlying
physical mechanism (which is...what, again?), an obsever who witnesses
a brightening of several magnitudes accompanying an ejection of
0.1--10 $M_{\odot}$ from a massive H-rich star would classify it as a
giant LBV eruption, because that's the definition of an LBV eruption.
So, whether one wishes to call them giant LBV eruptions, SN impostors,
LBV impostors, pulsational pair instability ejections, failed SNe,
explosive shell burning events, mergers, or some other name, the fact
remains that if seen in an external galaxy, we'd probably call it a
giant LBV eruption.

\section{Massive Stars Can Die With Hydrogen}

To me, one of the most interesting questions in massive star research
is whether these Type IIn SN progenitors (1) are stars that really
share the same evolutionary phase as local examples of LBVs but are
exploding as SNe, or instead (2) are caused by some different
underlying physical mechanism that causes LBV-like mass loss right
before the SN explosion.  There's also (3) the possibility that some
of them are genuine pair instability SNe (Smith et al.\ 2007b).  One
or more of these is right, but no matter which one it is, a firm
observational fact remains that is hard to avoid and which I want to
emphasize:

\smallskip

{\bf At nearly Solar metallicity, observed supernovae tell us that
very massive stars can make it to the ends of their lives and explode
with massive H envelopes still intact.}

\smallskip

This is a very important clue to understanding the evolution of
massive stars, because it is in direct conflict with the predictions
of stellar evolution models.  One might conjecture that the
predictions of stellar evolution models, which depend primarily on the
adopted mass-loss rates, are wrong because they have assumed mass-loss
rates that are too high...and we know they are too high.  If this is
true, then the current paradigm --- that LBVs represent only a very
brief transition phase between the end of core-H burning and the
beginning of core-He burning lasting a few 10$^4$ yr --- is probably
wrong as well, and the idea that LBVs can explode as SNe becomes more
compelling.

In order to make the normal LBV phase last until core collapse, the
LBV lifetime must be longer than current estimates of a few 10$^4$ yr
-- more like a few 10$^5$ yr --- because it must outlast core He
burning.  The main justification for a short LBV lifetime is that
there are too few of them, so statistically, their lifetimes can't be
too long.

Is it possible that the time during which an evolved massive star can
potentially be an LBV is actually longer?  Suppose, for the sake of
argument, that the specific eruptive instability that leads us to call
something an LBV actually represents an intermittent, possibly
recurring active phase within a much longer blue supergiant/LBV phase.
In other words, suppose LBV stars go through {\it dormant} phases,
like volcanoes, which last much longer than their eruptive phases.
What are the consequences of that?  There should be many more H-rich
evolved blue supergiants that are not caught at the right moment when
they show wild variability, but may or may not have observable
circumstellar material.  This is, of course, known to be the case.
There are many blue supergiants --- often called LBV ``candidates''
because of their spectral similarity to LBVs --- which do not exhibit
the specific variability that earns them {\it bona-fide} LBV status.
What are these stars, if not evolved massive stars that are potential
dormant LBVs?  Massey et al.\ (2007) argued a similar point, noting
that there are several hundred LBV candidates in M31 and M33, compared
to 8 known from their variability.  {\bf I would argue that this means
the ``greater LBV phase'' is in fact much longer than the very rapid
transition from core-H to core-He burning that we normally hear
quoted.}  If true, it would no longer be surprising to see LBVs
exploding as SNe.

\section{Addendum: Binaries, Binaries, Binaries....}

At this point, especially at this meeting, an obvious proclamation
comes to mind: ``Binaries are the solution!''  Namely, the requisite
LBV precursor event could occur in a companion star in a massive
binary system instead of the exploding star; the exploding star would
be a more evolved WN/WC star, so the Type Ib/c SN then expands into
its companion's dense H envelope, appearing as a Type IIn SN.  This
would nicely resurrect stellar evolution models, because then massive
stars don't need to survive until core collapse with H envelopes
intact.  Phew!

The problem is that this actually makes things much worse.  Remember
that the SN and LBV-like eruption need to be synchronized to within
about a decade to produce a Type IIn event.  {\it What are the chances
that the LBV star in a WR+LBV binary system would happen to have a
giant LBV eruption within a decade before the WR star explodes?}  At
this meeting, J.\ Eldridge noted that in binary evolutionary models
for massive stars, one expects that $\sim$2--5\% of WR stars will have
companions in the LBV phase.  But here I'd emphasize that having a
companion in the LBV phase is not enough for Type IIn SNe: we also
need to have that companion suffer a giant eruption $\lesssim$10 yr
before the other star's SN explosion.  How likely is that?  If the
nominal time between recurring giant LBV eruptions is 1000 yr, then
there's a 1\% chance of an LBV eruption occuring within a decade of
its companion's SN --- but that occurs only in the $\sim$5\% of
massive binaries already in the WR+LBV phase --- so now we are down to
0.05\%.  {\it This is not nearly enough.}  Type Ib/c SNe make up about
15\% of core-collapse SNe, while Type IIn's make up about 2--5\%
(Capallaro et al.\ 1997).  Therefore, we would need about 10--25\% of
the Type Ib/c SNe to explode into a companion's LBV-eruption envelope
to account for Type IIn SNe, compared to the 0.05\% we might expect.
What about mergers?  Given the compact radii of WR stars, it seems a
tall order to expect that $\sim$10\% of massive binaries would merge
fortuitously and eject a few solar masses of H only a decade before
the SN.



\begin{discussion}

\discuss{Davidson}{Nathan, you mentoned the word ``gonzo'' and
indicated that some astronomers imagine its negative!  In science,
{\it gonzo is good.}  Astronomy has only a few dozen gonzo objects,
typically with two attributes: (1) They're at extraordinarily
revealing stages in their careers, and (2) are close enough to observe
really well.  Apart from Eta Car, you mentioned two or three of the
others.  {\it Each is worth hundreds of routine objects}, because they
oft show where theory fails.  These objects often go negleced by
observers, sometimes at crucial stages.}

\end{discussion}

\end{document}